# Social Influences in Recommendation Systems

Diyah Puspitaningrum
*Department of Computer Science*
*University of Bengkulu*
Bengkulu, Indonesia
diyahpuspitaningrum@gmail.com

*Abstract*—Social networking sites such as Flickr and Facebook allow users to share content with family, friends, and interest groups. Also, tags can often assign to resources. In the previous research [4] using few association rules FAR, we have seen that high-quality and efficient association-based tag recommendation is possible, but the set-up that we considered was very generic and did not take social information into account. The proposed method in the paper [4], FAR, in particular, exhibited a favorable trade-off between recommendation quality and runtime. Unfortunately, recommendation quality is unlikely to be optimal because the algorithms are not aware of any social information that may be available. Two proposed approaches take a more social view on tag recommendation regarding the issue: *social contact variants* and *social groups of interest*. The user data is varied and used as a source of associations. The adoption of social contact variants has two approaches. The first social variant is User-centered Knowledge, to contrast Collective Knowledge. It improves tag recommendation by grouping historic tag data according to friend relationships and interests. The second variant is dubbed 'social batched personomy' and attempts to address both quality and scalability issues by processing queries in batches instead of individually, such as done in a conventional personomy approach. For the social group of interest, 'community batched personomy' is proposed to provide better accuracy groups of recommendation systems in contrast also to Collective Knowledge. By taking social information into account can enhance the performance of recommendation systems.

## I. Introduction

Two kinds of recommender systems currently dominate the literature. The first kind consists of recommender systems based on "collective knowledge," i.e., extensive collections of historic tagsets assigned to resources, by any users. When each user can assign his/her tagset to a resource, this is usually called a 'folksonomy.' It is contrary to the set-up that is considered in the previous research [4]. The second kind consists of 'personalized' recommender systems that only use annotations from the history of a given user. Such personomy generally performs well if the user is somewhat active and thus has sufficient history.

Hence, existing systems either assume that available tag data should be used or only personal tag data. However, in the case of social networks such as Flickr and Facebook, users can also maintain 'friendship' relations with other users. Since it seems safe to assume that their friends and family influence people, this suggests a third type of recommender system after collective context and personal context, i.e., one that exploits 'social' information. Hence, the question is the following: *Can social information improve the quality of tag recommendation systems?*

## II. Related Work

The generic association-based recommendation methods by Sigurbjornsson and Van Zwol [8] and Menezes et al. [6] have already discussed in [4]. This paper only discusses relevant work in the context of social tag recommendation. Rae, Sigurbjornsson, and van Zwol [7] proposed to combine different `contexts' to improve tag recommendation in social networking environments. They investigated whether the Social Group context (a user's interest group) and Social Contact Context (a user's friends) can contribute to the performance of the system, in particular when used in combination with the more conventional Personal and Collective context. They do this with a straightforward voting approach. As a baseline, they use both the Personal Context and the Collective Context. According to their paper, Social Contact Context (i.e., using a user's friends' tag data) perform so bad that it is harmful to overall performance when used in combination with other contexts. In other words, the tagging behavior of a user's contacts is unhelpful for tag recommendation.

Different from their approach [7], we explore the combination of different information sources by physically selecting/appending tag data, and then performing recommendations based on the newly created knowledge source. It is different from a voting approach, as data instead of associations are combined, and the resulting associations are thus appropriately weighted. Hu, Wang, Liu, and Li [2] investigated social influence by Gaussian-type topological potential with user influence ranking is ranked by social contact networks. As the output is the top-$N$ contacts to form a 'preference community.' Tagging information from friends-of-friends is used to make the recommendation more diverse, then through measurement of user social influence, they find those who have an impact on the target user [2].

For a social group of interest tag recommendation, in social media sharing sites such as Flickr, a user can subscribe to group(s). This kind of membership explicitly interests in some topics. Garg and Weber [1] proposed a personalized approach to tag recommendation by doing query dependent group tag usage based on the group where the photo belongs. Rae, Sigurbjornsson, and van Zwol [7] use the group context of the user where photos that have not been assigned yet to groups implement to it. A 'social group context' is obtained by aggregating the photo annotations of photos posted in the groups where the user subscribed. Their work is similar to the user-tag matrices called personomy used by Jaschke et al. [3]. The difference is Jaschke et al. using implicit tag usage similarity between users.

Different from other works mentioned above, applied in an explicitly social group of interest, data is made up based on the group the photo belongs to and user history, but using collective knowledge for users without membership of a group to take into account scalability by the batch process.

III. MATERIAL AND METHODS

*A. Problems*

**Problem 1** (Social Contact Tag Recommendation). *Given a source database **D** over tag vocabulary **T**, users **U** and corresponding social network **G**, and an input tagset **I** provided with user **u** recommend a set of tags R(D, G, I,u) ⊆ (T \ I) ranked by relevance.*

**Problem 2** (Social Group of Interest Tag Recommendation). *Assume given a source database **D** over tag vocabulary **T**, users **U** corresponding communities or groups of interest **C**, and an input tagset **I** provided with user **u** who recommend a set of tags R(D, C, I,u) ⊆ (T \ I) ranked by relevance.*

*B. The Methods*

*1) Tools*

In the previous research [4], we investigated the performance of the state-of-the-art when it comes to association-based tag recommendation, and proposed additional methods ourselves. In [4], we used three main association rules variants: PAIRAR (PAR), NAIVEAR (NAR), and FASTAR (FAR) as base methods, but we will vary the source database *D* as input for the recommendation methods. PAIRAR or PAR, for short, is the name of the method proposed by Sigurbjornsson and van Zwol [8]. NAIVEAR or NAR is a generalization that employs tag associations of *any length*, that is, up to length |*I*|+1 (which is sufficient to compute all needed conditional probabilities).

Not only NAR, LATRE, introduced by Menezes et al. [6], is a very similar method that also uses more extended associations. To overcome the pattern explosion, LATRE uses on-demand mining of association rules for each query. Three parameters can be used to reduce the number of rules: 1) a maximum number of tags per rule, 2) minimum support, and 3) minimum confidence. The main difference between LATRE and NAR is that NAR uses higher minimum support to mine associations once beforehand and augments these with the top-*m* co-occurring tags for each input tag.

In [4] we show that NAR results in the highest precisions, but also requires most runtime. The last, FASTAR (or FAR), which augments the pairwise associations from PAR with associations selected with KRIMP. In FAR, computing the scores for the candidate tags is done as by PAR and NAR, except that code table, *CT* consults for any conditional probabilities that one cannot obtain from the top-*m*. For these kinds of cases, $\hat{\text{sup}}_D(X)$ estimates the support of a tagset *X*. The code table itself computed offline. While computing recommendations, one only needs to estimate supports. Since a code table is smaller than the set of all frequent tagsets, compared to NAR, the FAR results in slightly lower precisions but much shorter runtimes. This paper uses only NAR and FAR due to their high precision than PAR. The FAR algorithm should be much faster than both NAR and LATRE.

*2) Social Perspective of Tag Recommendation*

*a) Social Contact Tag Recommendation*

In the previous research [4], we used *all* available historic tag data as a source of information for a recommendation, i.e., as *D*. We will refer to this choice of source database *D* as *collective knowledge* (*CK*), a term that has been used before [8]. When user (but no social) information is available, this is sometimes called a *folksonomy* [3]. At the other extreme, one could consider users one by one. To be more precise, all historic tag assignments of user u, denoted by $D_u$, can be used as the source database for the recommendation. This case is usually called a *personomy* [5]. Although such a set-up leads to fully personalized recommendations, this also has some disadvantages: 1) this approach cannot be used when no historic assignments are available for a user, and 2) making full-fledged personalized association-based recommendation can be computationally intensive, as associations need to be mined for each user individually. Also, observe that neither of the two approaches takes the social graph into account. In this research, we will take the social graph into account.

- **User-centered Knowledge**. This paper focuses on both a user's social network and his or her interests. In terms of the amount of tag data used, it lies somewhere between collective knowledge and a personomy. As has been argued before [7], several social factors play an essential role in improving recommendation quality. Firstly, friends influence a user; hence, using tags assigned by a user's friends can be beneficial. Obviously, to a lesser extent, the same can be said, such as for friends of friends, friends of friends of friends. However, friendship alone is not enough, as users may have very different friend groups (e.g., family, soccer friends, college friends). The situation leads to the second social factor: shared interests. Friends that share the same interests are more likely to use the same tags.

  This paper proposes to use *user-centered knowledge* (*UK*), consisting of tag data that is grouped by both friendship relations and user interests. That is when an association-based tag recommendation algorithm invokes a query *I* and user *u*, a source database *D* should use that consists of tag data 1) from all users within the *n*th-degree of *u*, 2) that concerns the same high-level interest as query *I*.

- **Social Batched personomy**. This paper already briefly mentioned personomy, being the historic tag data assigned by an individual user. To avoid the problem that no (or very little) historical data is available for many users (*cold-start*), one could consider exploiting the social graph by adding historical data from a user's friends. Unfortunately, this only resolves part of the problem, as the approach still needs to process each query/user individually. In an online environment, this may be undesirable or even infeasible.

  The *batched personomy* applies in large scale systems with 10s to 100s or even 1000s of given queries per second. Instead of processing each query individually, the process can also be held in batches, as follows.

  Step 1: the system does not process each query but waits until either a maximum number of queries or maximum waiting time reach.

  Step 2: the cached queries form a batch *B*.

Step 3: for a batch *B*, construct a database *DB* that contains previous tag assignments done by all users that have a query in *B*.

Step 4: if desired, add to *DB* historic tag data up to the nth-degree friends for each user.

Step 5: recommend tags for each query *I* in *B* using *DB* as the source database.

It is clear that such an approach is only possible when many queries come in at the same time, but considering the scale of current online social networks, this is entirely realistic. Given the right circumstances, the approach has potential advantages. First of all, by batching queries, associations have to be mined only once for each batch. Batching queries is much better than having to mine associations for each query, as is the case with personomy. Also, the source databases will be much smaller than when (all) collective knowledge is used, which may again result in a faster recommendation. Second, recommendation quality is likely to be better, on average. Batching queries is better than personomy because problem such as lack of historical data is smoothed out by forming batches. Also, batching queries is better than collective knowledge due to more personalized. The main difference between the user-centered knowledge approach is that groups are formed on-the-fly, rather than beforehand. Step 4 in the above procedure enables social information, but this is not required. When the step conducts, one can refer to this as a *social batched personomy*.

*b) Social Group of Interest Tag Recommendation (Community Batched Personomy)*

Groups are explicit expressions of members to create collections of topics they interest. Following is the introduction of the concept of *community batched personomy*, a partially processed batch. It is applicable in an extensive system to take benefit of classifying databases into the community and non-community databases. The non-community is processed using batch to speed-up execution time. The community data processed data per query by considering user history (*personomy*) and his online community, where he is a member. A user can belong to more than one community. In an extensive system, the batches can handle queries until, say, 1000s of queries per second. This approach is better than the one with a collective knowledge approach because the recommendation is more personalized and similar in database characteristics since they are talking in the same topic of interest. It is similar to asking queries to some groups of experts.

The community batched the personomy process as follows. First, the system classifies queries into two categories: from users who do not have groups and from users who have groups. Second, do the batched process for all users who do not have groups until the maximum capacity of the number of queries reach as follows. Cached queries form a testing set named batch *B'*. For a batch *B'*, construct a database *DB'* of collective knowledge for the training set. Recommend tags for each query *I'* in *B'* using *DB'* as a source database. Third, for users who have groups do per query processing: create *DB''* of historic tag data and group data; then recommend tags for each query *I''* using *DB''* as the source database.

IV. EXPERIMENTS

- **Evaluation Criteria**. To assess the recommendation quality, several metrics used as follows: precision (P@x), success (S@y), mean reciprocal rank (MRR), and time (milliseconds).

- **Implementation.** The base tag recommender algorithms PAR/NAR/FAR were implemented in C++ and called by the social group of interest program in Python for a recommendation. Experiments run on a Windows XP machine with an Intel Core i3 2.30 GHz CPU, 1.94GB RAM. The Flickr datasets took in January 2010 to early 2013.

*A. Social Contact Tag Recommendation*

- **Datasets**. To ensure each dataset satisfies the requirements of user-centered knowledge, first, a query for each of the two datasets, viz. 'London' and 'Paris' picked. Secondly, select a random user from the results of these queries. These gave one user with 'interest' London, and one with interest in Paris. From these two seed users, a breadth-first search to crawl tag data for all users within 4 degrees (or hops), constrained to tag data that contains the query. By strictly enforcing these queries, there is a shared high-level interest among the crawled groups of friends. From the quarter-million users in the beginning, after pre-processing, the dataset ended up with 1571 users, over 109680 photos, and a total of $|T|$=58930 different tags. London contains tag data for 54620 photos and Paris for 55060 photos. The largest tagset contains 73 tags. Finally, Combined is the combination of two datasets, London and Paris. Table I shows some basic properties of the three datasets. The '|t| (%)' denotes the percentage of transactions that have <6, 6-8, resp. >8 tags. '#friends' denotes the number of users that have X friends.

- **PAR, NAR, and FAR**. In all cases, use the top *m* = 50 most frequent co-occurring tags. For NAR, to obtain tagset collection ***F***, and mine all closed frequent tagsets with *minimum support* = 0.0007 and *maximum length* or *maxlen* = 3. The maxlen is uniform for simplicity. For FAR, CT induced with minimum support = 0.00007 and maxlen = 3.

- **Implementation**. The base tags: PAR, NAR, and FAR.

TABLE I. DATASET PROPERTIES

| Dataset | $|U|$ | $|T|$ | |t| (%) | | | #friends | | | | | |
|---|---|---|---|---|---|---|---|---|---|---|---|
| | | | <6 | 6-8 | >8 | 0 | 1-2 | 3-10 | 11-50 | 51-250 | ≥251 |
| London | 749 | 40106 | 21 | 24 | 55 | 655 | 0 | 0 | 2 | 30 | 62 |
| Paris | 874 | 33172 | 8 | 28 | 64 | 626 | 0 | 0 | 8 | 74 | 166 |
| Combined | 1571 | 58930 | 14 | 26 | 60 | 1243 | 0 | 0 | 10 | 102 | 216 |

*B. Social Group of Interest Tag Recommendation*

- **Datasets**. Datasets obtained from the Flickr photo-sharing (http://www.flickr.com/services/api/). Initially, crawled data by input query of 'Paris' to groups, and get members and photo tags per *k*-top list of groups in which the groups have enough database (*k* = 31). For users who have a small size database of groups, they similar to users without a group. All transactions consist of at least two tags. The data consists of 262135 photos over a total of |T|=58889 different tags. The minimum length of tags per transaction is 2, whereas the maximum length is 88 tags. Fifty-seven groups exist in the datasets, and the dataset splits by 5-fold cross-validation.

- **FAR**. In all cases, it uses the same maximum length for simplicity. A code table CT induced by KRIMP consists of closed itemsets with *maximum length* = 3 and *minimum support* = 0.0003.
- **Implementation.** The base tag recommender method is FAR.

## V. RESULTS

### A. Social Contact Tag Recommendation

The first series of experiments concerns the comparison of the user-centered and collective knowledge approaches. Table II shows the results obtained on the Combined Flickr dataset. The 5-fold cross-validation was used to create the train/test datasets. Each tagset in the Combined dataset is used once for testing, and the results average over all these results, but which data used as source database $D$ depends on the specific approach. In the case of User-centered Knowledge (UK), the training data restricts to either the London or Paris data, depending on the origin of the test tagset. These simulate using tag data that is within the $n$th degree of a user corresponding to the same interest.

For $k$=1, the results of NAR and FAR are equivalent to that of PAR; hence, this paper only presents the latter. The first --and the most important-- observation is about the independence of the used method: the UK approach *always* provides better results than CK. The situation implies that exploiting social and interest information can contribute to higher quality association-based recommendations.

The second observation is that given a $k$ and $D$, the three methods, PAR, NAR, and FAR, the NAR gives high precision while the runtimes are 2.25 to 20.71 times slower than PAR; the FAR achieves similar albeit slightly less accurate recommendation but is as about as fast as PAR. The FAR consistently gives a favorable trade-off between runtime and recommendation quality. These experiments show that this is also the case in the context of social tag recommendation, on a different type of tag data.

In the remaining experiments, an alternative method was used to generate the train/test data. The reason for this is that 5-fold cross-validation gives test sets that are too small for the more personalized approaches that being evaluated now. Consequently, one cannot compare the results in Table II to those in Table III, IV, and V. In the remaining experiments, for each of 5 'folds', randomly select 40% of the Combined dataset as test data, and use the remainder for training. All averages over 5 'folds'.

In the second series of experiments evaluates the batched personomy approach. For each fold, the entire test set is a single batch. Select corresponding data from the training set correspond to each of the users that have a query in the test set. This situation gives the default of the 'non-social' batched source database used for recommendation denoted as $D_{batched}^u$. Then, construct social-perspective variants by including up to $n$th-degree friends for each user $u$ that has a query in the batch, denoted $D_{batched}^n$, for $n$={1,2,3}. For example, $D_{batched}^1$ is equal to $D_{batched}^u$ augmented with tag data supplied by the direct friends of all users in the batch. When $D$ is empty after construction as just described, the method falls back to using the collective knowledge, to ensure that the system recommendation can give a recommendation. FAR is used as a recommendation algorithm (except when indicated otherwise), because it performed well in the previous experiment series. Results obtained in UK and CK with NAR are also available for comparison purposes.

Some interesting observations present in Table III. First, $D_{batched}^u$, the default (non-social) batched personomy consistently results in a higher P@1 than the collective knowledge approach. The opposite is true for P@3 and P@5, indicating that a personalized approach improves the recommendation of particular tags, but this may come at the cost of diversity. Second, however, is that exploiting social information can compensate for this: using user-centered knowledge always give the best performance in terms of precision. Third, is that one can trade-off runtime with precision by using social batched personomy, with either only direct friends, or friends and friends-of-friends. In the two rightmost columns shows percentages of the data used on average for a batch, and that using less of the data leads to faster runtime and lower precision.

TABLE II. RECOMMENDATION PERFORMANCE: USER-CENTERED AND COLLECTIVE KNOWLEDGE (UK RESP. CK), ON THE COMBINED DATASET

| k | Method | D | #tagsets | P@1 | P@3 | P@5 | S@3 | S@5 | MRR | time |
|---|--------|----|----|------|------|------|------|------|------|------|
| 1 | PAR | UK |  | 52.93 | 41.71 | 35.54 | 67.13 | 72.13 | 60.46 | 426 |
|   | PAR | CK |  | 47.35 | 38.14 | 32.87 | 62.98 | 69.39 | 55.76 | 568 |
| 2 | PAR | UK |  | 59.77 | 50.66 | 44.18 | 75.72 | 80.18 | 69.22 | 314 |
|   | PAR | CK |  | 55.83 | 46.93 | 41.00 | 72.24 | 77.60 | 65.79 | 527 |
|   | NAR | UK | 250822 | 60.73 | 51.39 | 44.84 | 76.05 | 80.42 | 68.67 | 962 |
|   | NAR | CK | 194516 | 56.75 | 47.61 | 41.59 | 72.55 | 77.85 | 65.04 | 881 |
|   | FAR | UK | 10036 | 60.06 | 50.87 | 44.33 | 76.51 | 80.88 | 68.53 | 305 |
|   | FAR | CK | 15636 | 56.45 | 47.25 | 41.24 | 73.05 | 78.34 | 65.13 | 514 |
| 3 | PAR | UK |  | 61.38 | 54.75 | 48.74 | 79.57 | 83.95 | 71.76 | 288 |
|   | PAR | CK |  | 58.89 | 51.07 | 45.40 | 76.35 | 81.48 | 69.19 | 589 |
|   | NAR | UK | 250822 | 62.92 | 55.33 | 49.19 | 79.57 | 83.86 | 71.51 | 4040 |
|   | NAR | CK | 194516 | 59.91 | 51.55 | 45.78 | 76.50 | 81.49 | 68.53 | 2169 |
|   | FAR | UK | 10036 | 62.20 | 54.76 | 48.81 | 80.67 | 84.93 | 71.60 | 258 |
|   | FAR | CK | 15636 | 59.92 | 51.42 | 45.68 | 77.73 | 82.67 | 69.09 | 463 |

Table IV shows the results for a subset of the same experiments in more detail, i.e., only for the London data with $k$=3. The London dataset is much more stringent than the Paris one, as the datasets are more or less of the same size, and the averages in Table III are much higher. The relative differences between the different approaches are pretty much the same, however. The reason is that they allow a comparison for the next experiment.

Given that the batched personomy method strongly inspired by the personomy approach, the question that naturally arises is how these two approaches compare. The experiment in which queries processed individually performs, and because this is computationally rather intensive, it only did for a single setting: for dataset London, with $k$=3 and NAR. The results presented in Table V. For $D^u$, it is only use training data provided by the user giving $u$, unless $u$ have no such history. In that case, the complete training set is used to ensure that the recommendation system can result in a recommendation. $D^1$, $D^2$, and $D^3$ are social personomy, completely analog to their batch counterparts.

TABLE III. BATCHED PERSONOMY ON THE COMBINED DATASET

| k | D | #tagsets | P@1 | P@3 | P@5 | MRR | time | $\frac{|D|}{|UK|}\%$ | $\frac{|D|}{|CK|}\%$ |
|---|---|---|---|---|---|---|---|---|---|
| 1 | $D^u_{batched}$ | 10097 | 36.45 | 26.22 | 20.92 | 43.93 | 77 | 23 | 12 |
|   | $D^1_{batched}$ | 8674 | 37.61 | 26.87 | 21.32 | 45.34 | 154 | 54 | 27 |
|   | $D^2_{batched}$ | 10138 | 39.08 | 28.16 | 22.38 | 47.09 | 251 | 85 | 43 |
|   | $D^3_{batched}$ | 11494 | 39.27 | 28.31 | 22.48 | 47.30 | 234 | 99 | 49 |
|   | UK | 11085 | 39.14 | 28.26 | 22.39 | 47.20 | 256 | 100 | 50 |
|   | CK | 15362 | 35.16 | 26.76 | 21.91 | 44.21 | 622 | 200 | 100 |
|   | UK + NAR | 225060 | 40.39 | 29.09 | 23.08 | 48.44 | 465 | 100 | 50 |
|   | CK + NAR | 181116 | 35.16 | 26.76 | 21.91 | 44.21 | 302 | 200 | 100 |
| 2 | $D^u_{batched}$ | 10097 | 42.02 | 31.38 | 25.65 | 50.93 | 75 | 23 | 12 |
|   | $D^1_{batched}$ | 8674 | 42.78 | 31.96 | 25.92 | 51.90 | 183 | 54 | 27 |
|   | $D^2_{batched}$ | 10138 | 44.04 | 33.37 | 27.2 | 53.50 | 187 | 85 | 43 |
|   | $D^3_{batched}$ | 11494 | 44.15 | 33.46 | 27.33 | 53.64 | 268 | 99 | 49 |
|   | UK | 11085 | 44.16 | 33.43 | 27.29 | 53.62 | 238 | 100 | 50 |
|   | CK | 15362 | 40.68 | 31.83 | 26.6 | 50.90 | 637 | 200 | 100 |
|   | UK + NAR | 225060 | 45.69 | 34.47 | 28.15 | 54.89 | 1201 | 100 | 50 |
|   | CK + NAR | 181116 | 41.24 | 32.02 | 26.71 | 51.15 | 451 | 200 | 100 |
| 3 | $D^u_{batched}$ | 10097 | 43.81 | 33.13 | 27.63 | 53.12 | 71 | 23 | 12 |
|   | $D^1_{batched}$ | 8674 | 44.14 | 33.65 | 27.96 | 53.84 | 138 | 54 | 27 |
|   | $D^2_{batched}$ | 10138 | 45.67 | 35.24 | 29.35 | 55.61 | 179 | 85 | 43 |
|   | $D^3_{batched}$ | 11494 | 45.63 | 35.32 | 29.45 | 55.68 | 265 | 99 | 49 |
|   | UK | 11085 | 45.64 | 35.29 | 29.41 | 55.68 | 262 | 100 | 50 |
|   | CK | 15362 | 42.44 | 33.87 | 28.79 | 53.25 | 661 | 200 | 100 |
|   | UK + NAR | 225060 | 47 | 36.45 | 30.28 | 56.76 | 5796 | 100 | 50 |
|   | CK + NAR | 181116 | 42.96 | 34.07 | 28.84 | 53.39 | 906 | 200 | 100 |

TABLE IV. BATCHED PERSONOMY RESULTS FOR 'LONDON' (K=3, FAR)

| D | #tagsets | P@1 | P@3 | P@5 | MRR | time | $\frac{|D|}{|UK|}\%$ | $\frac{|D|}{|CK|}\%$ |
|---|---|---|---|---|---|---|---|---|
| $D^u_{batched}$ | 10893 | 34.98 | 27.47 | 23.19 | 44.52 | 85 | 22.65 | 11.28 |
| $D^1_{batched}$ | 8368 | 34.71 | 27.12 | 22.82 | 44.31 | 123 | 34.49 | 17.17 |
| $D^2_{batched}$ | 9527 | 37.53 | 29.91 | 25.27 | 47.45 | 181 | 71.02 | 35.37 |
| $D^3_{batched}$ | 12138 | 37.54 | 30.07 | 25.46 | 47.63 | 316 | 97.75 | 48.68 |
| UK | 11328 | 37.51 | 30.02 | 25.39 | 47.59 | 265 | 100 | 49.8 |
| CK | 15362 | 33.45 | 28.5 | 25.18 | 44.79 | 609 | 200.81 | 100 |

TABLE V. PERSONOMY RESULTS FOR 'LONDON' WITH K=3 AND NAR

| D | #tagsets | P@1 | P@3 | P@5 | MRR | time | $\frac{|D|}{|UK|}\%$ | $\frac{|D|}{|CK|}\%$ |
|---|---|---|---|---|---|---|---|---|
| $D^u$ | 22787756 | 43.58 | 38.02 | 34.23 | 48.53 | 30102 | 27.6 | 13.75 |
| $D^1$ | 25817746 | 42.98 | 37.25 | 33.51 | 47.84 | 30092 | 26.51 | 13.2 |
| $D^2$ | 27347812 | 42.84 | 37.17 | 33.47 | 47.75 | 31176 | 26.94 | 13.42 |
| $D^3$ | 26304926 | 43.06 | 37.25 | 33.52 | 47.96 | 30965 | 26.5 | 13.2 |
| UK | 204114 | 39.3 | 31.43 | 26.49 | 49.12 | 2142 | 100 | 49.80 |
| CK | 181116 | 33.69 | 28.85 | 25.32 | 44.83 | 922 | 200.81 | 100 |

The most important observation to be made concerns the total number of tagsets that have been mined to be able to provide the requested recommendations. In this query by query set-up, the number of mined tagsets by large exceeds all numbers so far. As a logical side-effect, the average runtime per query also increases, to about 30 seconds per query. This situation is way too long for an online environment. On the upside, this non-batched personomy approach leads to much better recommendation than its batched counterpart, as can be observed by comparing their obtained precisions in Table IV resp. V.

### B. Social Group of Interest Tag Recommendation

Table VI shows the *community batched personomy*. The precision value of recommendation using *community batched personomy* improves with a range between 13% to 20% compare to collective knowledge.

TABLE VI. COMMUNITY BATCHED PERSONOMY FOR 'PARIS' WITH K={1,2,3} AND FAR WITH MAXLEN[3]-0.0003D

| D | #input | #tagsets | P@1 | P@3 | P@5 | MRR | Time | $\frac{|D|}{|CK|}\%$ |
|---|---|---|---|---|---|---|---|---|
| GK | 1 | 1983 | 0.53 | 0.36 | 0.29 | 0.61 | 386.1 | 28.8 |
| CK | 1 | 6876 | 0.47 | 0.32 | 0.25 | 0.56 | 118 | 100 |
| GK | 2 | 1983 | 0.55 | 0.41 | 0.32 | 0.65 | 387.1 | 28.8 |
| CK | 2 | 6876 | 0.48 | 0.33 | 0.26 | 0.64 | 117 | 100 |
| GK | 3 | 1983 | 0.57 | 0.4 | 0.33 | 0.68 | 509.3 | 28.8 |
| CK | 3 | 6876 | 0.47 | 0.33 | 0.26 | 0.58 | 95.4 | 100 |

GK : Community batched personomy (group knowledge)    CK : Collective knowledge

## VI. ANALYSIS

Whereas friendship relations between users in a social network are readily available and therefore straightforward to exploit, this is on itself not enough to guarantee high-quality recommendation [7]. Friendship relations are only useful when they shared an interest that matches the current query. This situation leads to propose the User-centered Knowledge approach, which combines these two requirements.

Translating the high-level requirements for such a user-centered knowledge approach to a principled methodology is, unfortunately, far from trivial. Using tag data for friends up to the *n*th degree is easy, but ensuring that only friends corresponding to the current 'interest' are selected is not. This paper is a proof of the concept of simulating two user-centered datasets based on queries and friendship relations. With the uprising of community detection methods that take user interests into account, more principled and practical solutions will probably be available in the shortcoming future. This situation is promising because the experiments clearly showed that taking friendship and interest information into account leads to a better recommendation. It is about the recommendation quality and the runtime.

As for the personomy approaches, the query-by-query approach is infeasible for the association-based recommendation. Recommendation accuracy is high, and the method might be useful for offline situations, but runtimes are incredibly high. Even if all the algorithms used are optimized, the difference between the CK/UK/batched personomy approach and the personomy approach would remain significant. The social batched personomy approach appears to be a viable alternative. The friendship degree varies to trade-off runtime with recommendation quality. Further experiments are needed to investigate how this approach performs in the more general case when the data is not composed of two coherent parts.

The performance of the three base methods, PAR, NAR, and FAR, has proven consistent in the social context. As before, FAR provides a desirable trade-off between runtime and recommendation quality, and remains the proposed algorithm of choice when it comes to association-based tag recommendation. It is flexible enough to take different sources as inputs and still provide excellent recommendations.

For the social group of interest tag recommendation, as an online community is readily available to be exploited, it is proven that community batched personomy based on FAR is one of the approaches to mining-specific group(s) recommendation that offers a high-quality recommendation. It has shown that a reduced number of tagset in the code table leads to the faster recommendation, whereas proper selections of the group database maintain high accuracy.

## VII. CONCLUSIONS

The two proposed approaches take a more social view on tag recommendation. Instead of modifying the base algorithms, the varied data used for an association source. The first social variant, the User-centered Knowledge, is to contrast the Collective Knowledge. It aims to group historic tag data according to friend relationships and interests. The second variant is dubbed Social Batched personomy and processes queries in batches instead of individually. The empirical evaluation showed that tag recommendation could improve a recommendation system performance by taking social information into account.

For a social group of interest tag recommendation, the social tag recommendation exploits a specific community or *group of interest*. The approach aims to group historic tag data according to members of the community. The advantage of processing queries in batches is less operation time, while per query processing maintains a high quality of recommendation. The system could be improved by either taking group information or social information into account. For future work, a more difficult task is finding a way to detect membership when the community is implicit. Users are very likely to have several hobbies or interests, and members of more than one online community. The clustering algorithm then can be used to measure the distance among user clusters. After identifying a set of communities, one can use the set of communities for many purposes, such as enhanced browsing, identifying experts, website recommendations.